# Comparison of Decision Tree Based Classification Strategies to Detect External Chemical Stimuli from Raw and Filtered Plant Electrical Response

Shre Kumar Chatterjee[1], Saptarshi Das[1], Koushik Maharatna[1], Elisa Masi[2], Luisa Santopolo[2], Ilaria Colzi[2], Stefano Mancuso[2] and Andrea Vitaletti[3,4]

[1] *School of Electronics and Computer Science, University of Southampton, Southampton SO17 1BJ, UK*

[2] *Department of Agri-food Production and Environmental Science, University of Florence, Florence, Italy*

[3] *DIAG, Sapienza University of Rome, via Ariosto 25, 00185 Rome, Italy*

Email: s.das@soton.ac.uk, sd2a11@ecs.soton.ac.uk (S. Das*)

**Phone:** +44(0)7448572598, **Fax:** 02380 593045

## Abstract

Plants monitor their surrounding environment and control their physiological functions by producing an electrical response. We recorded electrical signals from different plants by exposing them to Sodium Chloride (NaCl), Ozone ($O_3$) and Sulfuric Acid ($H_2SO_4$) under laboratory conditions. After applying pre-processing techniques such as filtering and drift removal, we extracted few statistical features from the acquired plant electrical signals. Using these features, combined with different classification algorithms, we used a decision tree based multi-class classification strategy to identify the three different external chemical stimuli. We here present our exploration to obtain the optimum set of ranked feature and classifier combination that can separate a particular chemical stimulus from the incoming stream of plant electrical signals. The paper also reports an exhaustive comparison of similar feature based classification using the filtered and the raw plant signals, containing the high frequency stochastic part and also the low frequency trends present in it, as two different cases for feature extraction. The work, presented in this paper opens up new possibilities for using plant electrical signals to monitor and detect other environmental stimuli apart from NaCl, $O_3$ and $H_2SO_4$ in future.

## I.    Introduction

Plants, such as *Mimosa pudica* (Touch-me-not) and *Helianthus annuus* (Sunflower), show some form of physical changes due to external stimuli in the form of touch and sunlight respectively [1]. The wilting of general plants due to dry environmental conditions is also commonly found. For many years, researchers have tried to establish the relationship between these reactions of the plants and the surrounding environmental conditions [1]. It has been found that the underlying phenomenon behind this is the plant electrophysiological mechanism which may be traced in the electrical response of the plant to the external stimulus [1]. These electrical signals, which control various physiological functions in the plants, hold useful information about the external stimulus (which causes the electrical signal in the plant) contained within its deterministic and stochastic parts to different extents. Analysis using low frequency (trend) part of the plant electrical signal to study the external chemical or light stimulus has been reported in Chatterjee *et al.* [2], [3]. Also, other studies on plant electrical signal processing have been reported in [4]–[14], in particular use of classification techniques to find out the applied external stimuli, through various statistical features computed from the recorded plant electrical signal, was reported first in [2]. Since the statistical features in [2] were extracted from raw plant signals (with low frequency trends or drifts), a background (pre-stimulus) information subtraction method was adopted in the classification process to focus only on the incremental values in each feature due to the application of the stimulus.

In this paper, we initially focus on the information contained in the stochastic part of the plant electrical signals by applying a high pass filtering on the raw signals to remove the inconsistent trends or drifts. We also used raw signals with the trends (using the background information subtraction method as reported in [2]) to show a comparative analysis between the classification performance of the filtered and raw plant signals. Thus, we here explore, if there is any improvement in the classification process while using only the detrended random part rather than the raw signal containing small local fluctuations superimposed on relatively larger change in the trends.

In order to develop a classification strategy for detecting the external chemical stimuli, here we used 15 features out of which 11 features have been reported in [2]. In addition to these 11 features, four additional statistical features have been explored along with independent testing of the classifiers with a much larger dataset. In this paper, we report the use of *discriminant analysis* and *Mahalanobis distance* based classification algorithms to establish a decision tree based classification system using both *One-Versus-One* (OVO) and *One-Versus-Rest* (OVR) strategy [15], [16]. We also report the validation scheme of the classifiers in two different ways – 1) Leave One Out Cross Validation (LOOCV) on ~73% of the available data (*retrospective study*) and 2) independent testing of the remaining ~27% of the data (*prospective study*).

Datasets from experiments on plants using three different stimuli – NaCl, $H_2SO_4$ and $O_3$ have been used in this exploration and can be found on the EU FP7 funded project PLants

Employed As SEnsor Devices (PLEASED) website. The work presented in this paper may help in taking a step closer to the concept of a plant electrical signal based external stimuli sensing platform which the PLEASED project aims to develop. Such a device, if successful, will aid in monitoring a large geographical area for multiple environmental stimuli or pollutants of interest. In order to proceed towards the realization of such a stimulus classification scheme using plant electrical signals, the steps shown in Figure 1 were followed.

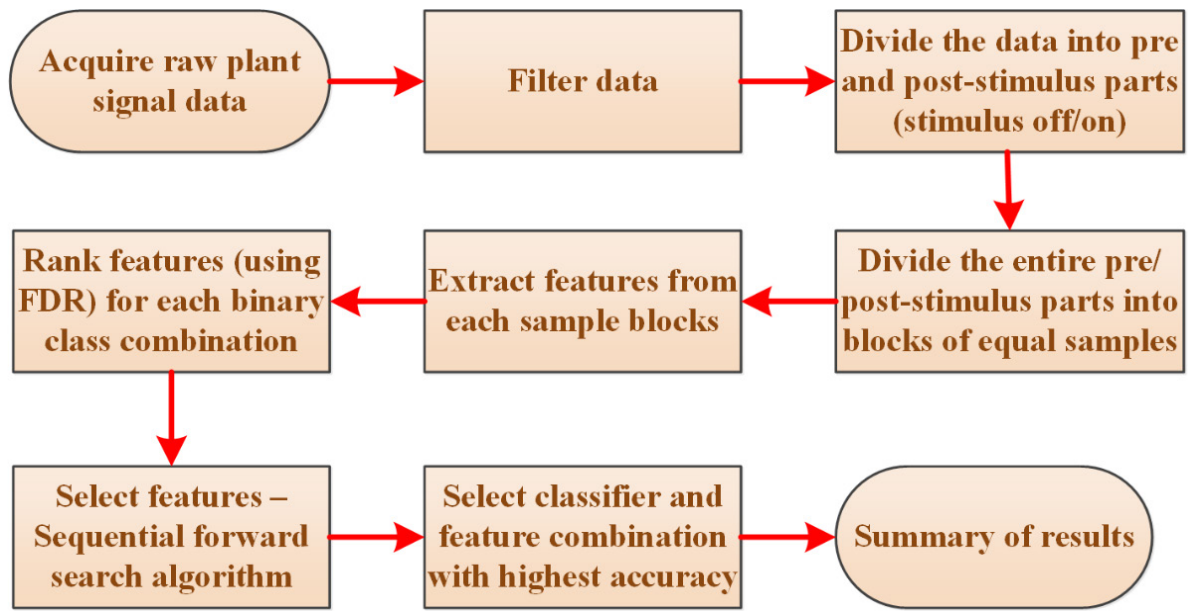

Figure 1: Steps for classification of environmental stimuli using plant electrical signal.

The work presented in this paper presents the following salient contributions over previous approaches:

- The present work uses filtered (containing only stochastic part of the signals) as well as raw plant signal (containing both deterministic and stochastic parts of the signals) for classifying the external stimuli for a comparative analysis between two pre-processing techniques affecting the final classification results.

- An exhaustive set of experimental data with 28,070 data blocks were used for training the classifiers, which is ~7.4 times higher than the data used in our previous work [2].

- Apart from Tomato and Cucumber plants which were used for the experiments reported in [2], Cabbage was included in the present work for extracting the experimental data. This helped to analyse the electrical response in a pool of three different species due to chemical stimulus.

- Features from NaCl of different concentration (5ml and 10 ml) were combined to be labelled as NaCl in the present classification work, whereas in the previous work [2], these two stimuli were considered as separate classes.

- In the previous study [2], the classification results were obtained using individual (univariate analysis) or feature pairs (bivariate analysis), whereas multivariate feature analysis have been carried out here.

- Apart from the previously explored 11 statistical features in [2], here we explore four additional features of the plant electrical signal for classification of the external stimuli *viz.* higher order central moments (hyperskewness and hyperflatness), fano factor, and correlation dimension.

- The classification results reported in the previous study [2] was average results for six binary stimuli combinations that in a way averages the detectability of one class with the others. Whereas a systematic decision tree has been developed in the present study that helps answering the question which chemical stimulus can be easily detected from the plant signals and which stimuli are hard to differentiate from the rest.

- Along with reporting the retrospective classification accuracy employing LOOCV, a separate held out dataset was used for prospective validation in this work, whereas only retrospective LOOCV results were reported in the previous work [2], on a much smaller dataset.

## II.  Recording Electrical Signal from Plants

Raw electrical signals from different experiments involving $O_3$, NaCl and $H_2SO_4$ as external stimuli, under laboratory conditions, were acquired from different plants. Each experiment was conducted on a new plant, thereby eliminating the risk of any residual effect of the previous experiments infiltrating the current electrophysiological condition of the plants. The experiments were conducted inside a plastic transparent box placed in a Faraday cage kept in a dark room, as shown in Figure 2, to minimize any external electrical or light interference. Artificial lights using LED lamps were provided in a 12 hours darkness / 12 hours light cycle to cater for the plant's photosynthetic needs. The plastic enclosure was used to introduce ozone and this enclosure being transparent is important so that the plant gets light for photosynthesis under normal condition. During chemical stress, the room was made dark as light can generate a different type of action potential thus masking the effect of chemical stimuli. For each experiment, three stainless steel electrodes from Bionen S.A.S, each of 0.35 mm in diameter and 15 mm in length, were used. The polarization effect of using steel electrodes may be studied in the future which is not considered here. These electrodes, similar to those used for Electromyography (EMG), were inserted in the middle and at the top of the stem of the plant for data acquisition and in the base of the stem for reference. The other end of the electrodes was connected into a two channel high impedance electrometer (DUO 773, WPI, USA). Data recording was carried out at a rate of 10 samples/sec through a data acquisition system, provided by Labtrax, WPI. More detailed information on the experimental setup can be found in [2].

For introducing $O_3$ as a stimulus, an inlet silicone tube was placed in the plastic transparent enclosure, while a second outlet tube throws the $O_3$ from the box to a chemical hood. $O_3$ was produced using a commercial ozone generator (mod. STERIL, OZONIS) and the concentration inside the box was measured using a suitable sensor. Similarly, regarding $H_2SO_4$ or NaCl stimulus, a silicone tube inserted into the plant soil and connected to a syringe placed outside the Faraday cage for injecting the solution. Figure 2 shows the description of the experimental setup to administer the stimuli. Before applying the stimulus, a period of 45 minutes was allowed to the plant to recover after the insertion of the electrodes into the stem.

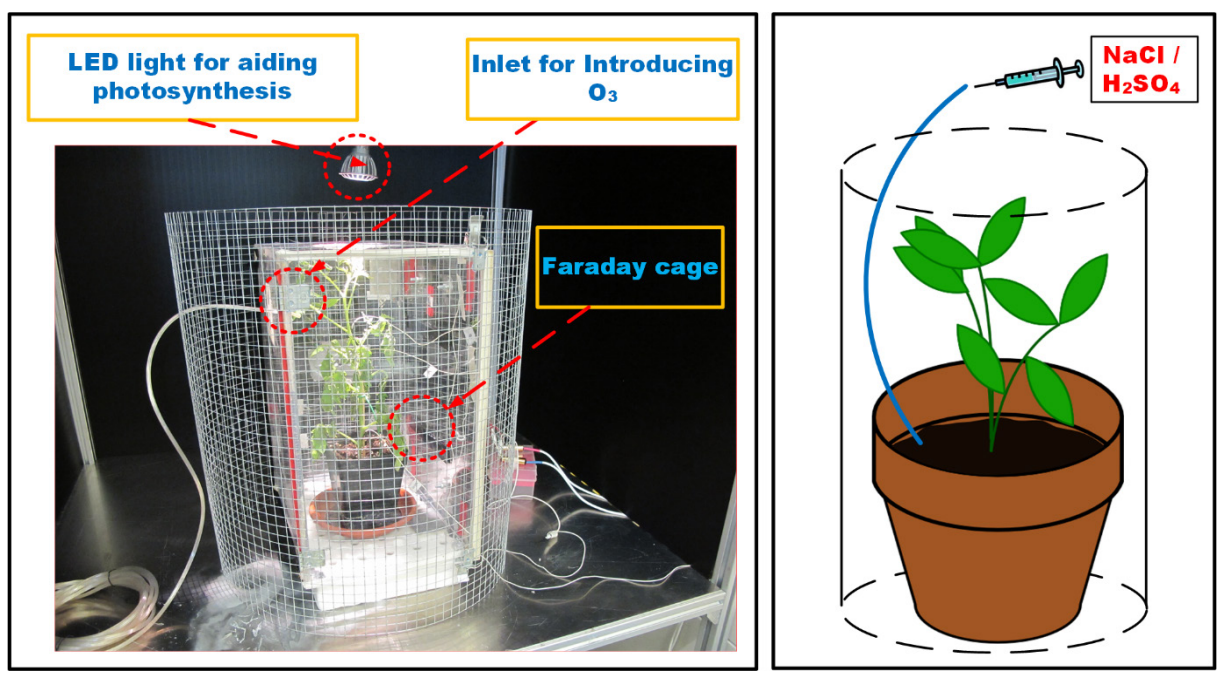

Figure 2: Experimental setup to extract plant electrical signal using different external stimuli.

Table 1: Details of the experiments with different chemical stimuli

| Stimulus | Concentration | Stimulus applied during each experiment | Plant species used | Number of independent time series |
|---|---|---|---|---|
| Ozone ($O_3$) | 16 ppm/ 13.07 ppm | Multiple | Tomato | 8 |
| | | | Cucumber | 4 |
| | | | Cabbage | 20 |
| Sulfuric acid ($H_2SO_4$) | 5 ml of 0.05/ 0.025 M solution | Once/twice | Tomato | 8 |
| | | | Cabbage | 20 |
| Sodium Chloride (NaCl) | 5/10ml of 3M solution | Once | Tomato | 16 |

Table 1 gives details about the experiments conducted to extract the datasets which has been used for the exploration presented in this paper. For experiments with $H_2SO_4$, the stimulus was applied once or twice to the plant in the form of 5 ml solution of 0.025M or 0.05M concentration of $H_2SO_4$. For experiments with $O_3$, the gas was injected into the box for 1 min every hour and the maximum concentration ranged between 13-16 ppm. NaCl treatment was carried out through addition of 5 or 10 ml of 3M NaCl solution to the soil. In [2], experiments using 5ml and 10ml of NaCl solutions were used as two separate classes. In the present work, we have combined both these concentrations of stimulus as a single class (i.e. NaCl), as here we want to explore the effect of NaCl as a whole while neglecting the slight variation due to its different concentrations.

Also, as is evident from Table 1, the experiments were conducted on multiple species of plants - Tomato (*Solanum lycopersicum*), Cucumber (*Cucumis sativus*) and Cabbage (*Brassica oleracea*) so that we can build a generalized classifier which is not specific to a particular plant species but is able to pick up the common signature in the electrical response of various different species. Whether some plant species are more sensitive to certain type of stimulus, can be explored in a future study. Therefore in this paper, we aim in making a generalized and robust detection technique for three chemical stimuli - $H_2SO_4$, NaCl, and $O_3$, rather than focusing on quantification of its variability due to the different species and concentration of the stimuli.

## III.    Pre-processing and Data Segmentation

### A.  Designing Optimum Filter for Drift Removal

In general it is not known which range of frequencies of the plant electrical signal contains useful information about the external stimulus which can be fed to the classifier. Therefore the first aim was to calibrate the pre-stimulus parts (due to different amplitude level for each experiment being different at the onset) and then bringing them at a common platform, so that the change in the post-stimulus part, upon application of a stimulus, could be quantified. We tuned the parameters of four digital high-pass Infinite Impulse Response (IIR) filters – Butterworth, Chebyshev type I, Chebyshev type II, and Elliptic filter [17], such that it minimizes the distance between the centroids of the clusters, formed using the distribution of wavelet packet energy along different wavelet basis for the pre-stimulus signals. We selected IIR over FIR filter structures due to lesser number of parameters which needed to be tuned. The objective function selected for the optimization is the energy contents of different leaf nodes in *wavelet packet decomposition* for the pre-stimulus parts of the plant signals which were acquired under laboratory settings, details of which have been reported in [17]. The IIR filter parameters were tuned in such a way that it produced almost overlapping clusters in the domain of wavelet packet energy along different wavelet basis for the pre-stimulus signals in different experiments, without enforcing the same for the post-stimulus signals (as this may mask the natural stimulated response of the plant). Therefore under such tuning of the filter as a pre-processing step for benchmarking of the background information (before any stimulus was applied), the post stimulus response can now be analysed to investigate any gross change in the statistical characteristics of the plant electrical signals. Through the exploration reported in [17], it has been found that a 6th order Chebyshev type-II high-pass filter yields the minimum cost function with a cut-off frequency of 0.77 Hz and stop-band ripple of 100 dB. Due to the fact that plant signals are slow in nature [1] and the bandwidth of the data is already constrained due to the present 10 Hz sampling rate, the recorded signals did not need to be low-pass filtered for the high frequency noise attenuation which might have come due to a faster sampling rate. However plant signals has significant drift in the low frequency part of the spectrum, removal of which has been addressed in [17]. In Figure 3, the top panels show example of raw signals, the middle panels show the high-pass filtered signals and the bottom panels show the Welch power spectrum estimate for the post-stimulus part. The dotted vertical lines indicate the time instant for application of the stimulus.

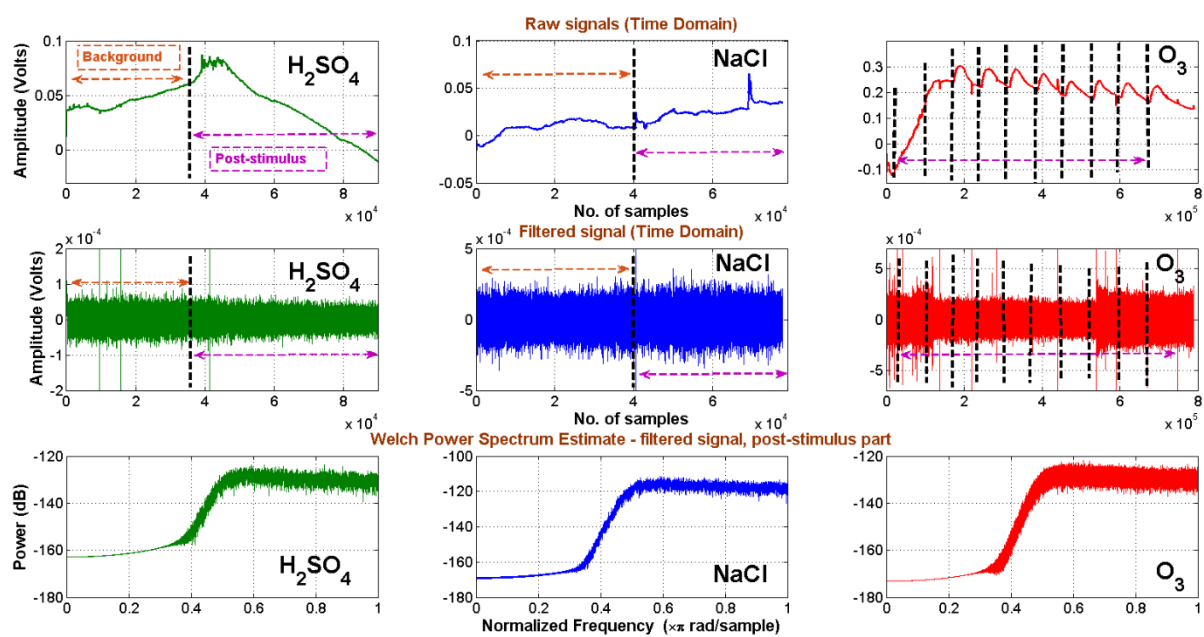

Figure 3: Plant electrical signals due to three different chemical stimuli.

Representative examples of the raw and filtered plant signals using this optimum filter are shown in Figure 3 for the three different chemical stimuli under consideration. The high-pass filtering removes the inconsistent low frequency drifts as evident from the almost horizontal detrended mean and the drop in power at low frequency in the Welch power spectrum estimates for the respective post-stimulus parts. It is worth mentioning that in most cases the drift is present in recorded plant electrical signals, like many other biological signals, due to variation in the instrumentation condition, environmental artefacts and changes in biological conditions over time. The possible reasons, effect and removal of the drift is discussed in a detailed manner in [17]. Due to the presence of the drift, either the signal has to be filtered to minimise its effect or the background information embedded in the features needs to be removed. We here compare both of these methods that yields better classification results.

From Figure 3, apparently it may appear that the power spectrums of the filtered plant signals under three different stimuli look quite similar (except few glitches in certain frequencies). This is in agreement with the findings in [2] and this study later reported that a simple spectral power as a feature cannot distinguish between these signals and thereby not provide good classification performance. Although the power spectrums look similar, there may be different information embedded within the plant electrical signals. These information's can be captured using higher order moments and other nonlinear statistical features which may not be captured with spectral power alone.

## B. **Data Windowing and Validation Scheme for the Classification**

Our previous work [2], [3] shows that plant electrical signals are slow in nature, therefore we here hypothesize that there will be sufficient information about the external stimulus within ~1.5 min of data segment (with sampling rate = 10 samples/second). In [17], a segment size of 256, 512, 1024 samples were chosen which are equivalent to ~0.42, 0.85, 1.7 min of streaming data respectively, in order to show the spectral energy distribution of the plant signals. The Fourier domain representation of the autocorrelation function i.e. the power spectral densities are explored in [17] and also in Figure 3 which show that the unfiltered signals have significantly high power in the low frequency due to the drifts which are reduced due to the optimum high-pass filtering. A very large sample size selection may introduce slow oscillations of the plant signal, thus corrupting the extracted features. It is important to note that after the optimum filtering only the high frequency components remain dominant as shown in Figure 3 bottom panels, denoting a fast decaying autocorrelation sequence.

Table 2: Blocks (of 1024 samples) for each stimulus in different validation schemes of the classifiers

| Validation Scheme | NaCl | $H_2SO_4$ | $O_3$ |
|---|---|---|---|
| Retrospective study (LOOCV) | 352 | 1340 | 26378 |
| Prospective study (independent testing) | 276 | 148 | 9692 |

Next, we divided the entire filtered signal into blocks of 1024 samples. In total $\approx 3.8 \times 10^4$ of such blocks were obtained (each block containing 1024 samples) from the post stimulus parts of different experimental datasets. Out of these total blocks, $\sim 10^4$ blocks (~27% of the total data blocks) were set aside for *prospective study* i.e. independent testing of the classifier settings. These data blocks for prospective study were obtained from completely separate experiments which were never seen by the classifiers.

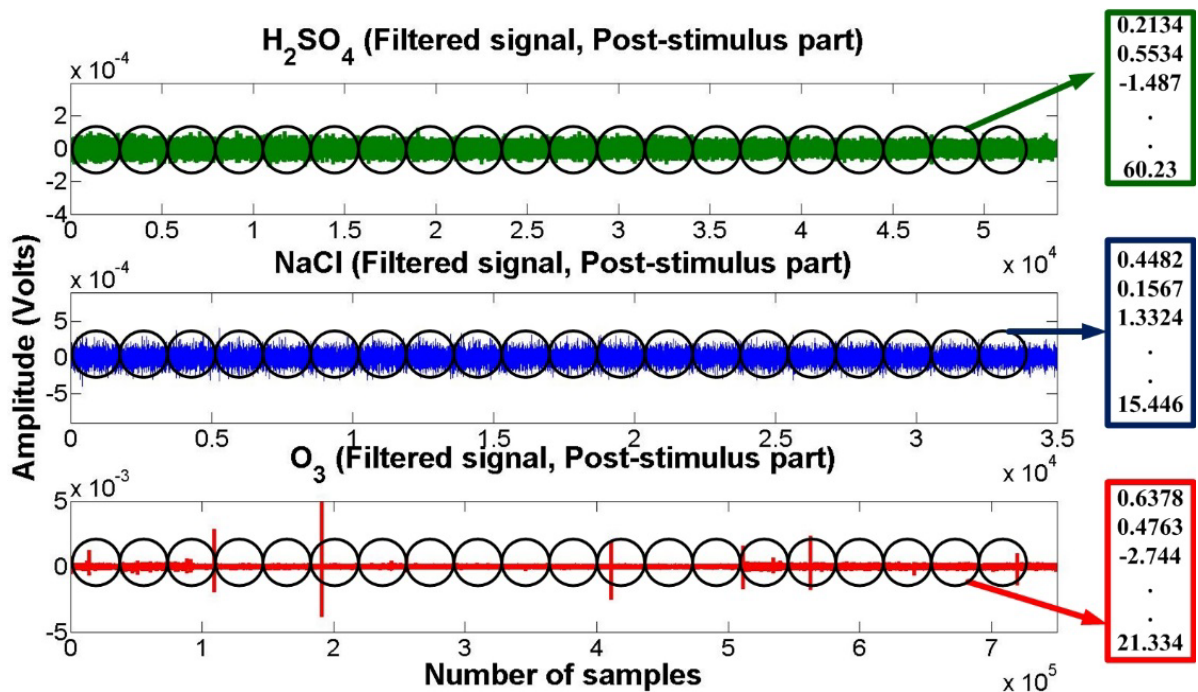

Figure 4: Dividing the filtered plant electrical signal into blocks of 1024 samples (shown by black circles) and computing feature values from each of these blocks (shown by red/green/blue boxes).

The data segmentation with non-overlapping window is shown in Figure 4 for the three stimuli, where the black circles refer to the blocks of 1024 samples from which the feature values are computed. The number of blocks used for each stimuli are shown in Table 2 from which we can see the main imbalance in the signal length was caused due to Ozone as a class, which contributed more data blocks due to the longer duration of the experimental recording than NaCl and $H_2SO_4$ [2].

The best classifier and feature combinations was selected by employing the LOOCV [18] on ~73% of the total data. The reason behind using only 73% of the data to find out the best feature-classifier combination were

a) to avoid selecting an over-fitted model of the total data, and

b) test the obtained best classifier-feature combinations on a section of independent data i.e. to estimate the performance of this combination in classifying completely unseen data.

## C. Rationale of segmentation of long plant signals for feature extraction and classification

In an ideal scenario of cheap and easy experimental data collection, it is often argued to train the classifier on a portion of the available dataset and then testing the model on a completely different experimental data commonly known as the held-out validation. Whereas we here employ a different approach, also known as the cross-validation scheme, commonly used when the independent realizations are small due to experimental constraints/costs involved. The problem with the held-out testing (particularly for smaller number of independent dataset) is that the results may introduce some bias which is typical to that particular segment of the held out dataset. In such a scenario, a smaller segment of the held out data in the testing phase may not always represent the common statistical properties of the whole dataset. Whereas a cross-validation reports average classification accuracy over all the segments of the data and is a much more robust measure of accuracy for judging the classification performance while working with small number of independent experimental datasets. Due to such scarcity of independent time series data coming from different experiments (as shown in Table 1), we first use resampling on the limited number of long recorded signals by windowing them in smaller segments. This resampling approach increases the number of data points in each class, so that a portion of the data can be held out to test the standard way of cross-validation in training phase, followed by independent held-out validation in testing phase for judging the classifier's performance. While working with limited number of independent experimental dataset, this resampling and data segmentation method has been widely used before and it is based on the assumption that the segmentation and resampling in the same long time series has similar statistical characteristics that we want to uncover using a classifier. This approach also assumes that different data segments coming from the same or limited number of long time series capture the effect of a particular stimulus in a similar way so that they can be grouped together using a classifier.

Huerta *et al.* [19] and Vembu *et al.* [20] reported dynamic support vector machine (SVM) with linear autoregressive (AR) kernel in similar time series classification problems. This is a viable option for time series classification, if the underlying data generation process is almost linear and smooth nonlinear. But for highly nonlinear and nonstationary time series, meaningful feature extraction that can capture these complex behaviours, becomes a necessity. For example in our study, the Hurst exponent, detrended fluctuation analysis (DFA) signify the long-range correlations and power law characteristics in the data, higher moments (skewness, kurtosis, hyperskewness, hyperflatness) signifying different forms of non-Gaussian attributes, correlation dimension and entropy signifying the extent of superimposed deterministic chaotic and stochastic noisy attributes in the data etc. [21].

The above mentioned method of using held out data may be a valid approach when the number of time series are significantly large where each independent time series can be represented as a single data point. In the case of relatively less number of available independent experimental dataset, a resampling statistics is a viable option, wherein the same long time series are chopped in several smaller segments to increase the number of data points. This resampling approach considers the data in each segment being independent and identically distributed (IID) and consider correlation only within the segment. Dividing long time series signals (with fewer independent realisations) in to smaller segments ignores the correlation lost between successive segments. In other words, it is assumed that the auto-correlated samples lie within a short range i.e. within the segment of 1024 samples but there is negligible autocorrelation between samples far apart, exceeding the window length. The choice of the window length and its effect on having a uniform background to compare the stimulated response has been discussed in [17]. Here, as a pre-processing step, we carried out signal detrending using the optimum high pass filter in [17] which significantly reduces the long distance autocorrelation of the samples so that a shorter window based feature extraction and classification can be applied effectively. This is a valid assumption for practical time series classification problem if the goal is to detect the cause behind the change in time series patterns by only looking at some features of the smaller segments. Here the segment size was chosen as 1024 samples (equivalent to 1.7 min) to extract the features. For any practical detection, this is a sufficiently long window for monitoring the chemical stimuli in the environment, rather than extracting features on the whole duration of the time series. This approach of resampling long time series in smaller segments has been previously used in many other biological time series analysis problems e.g. Electrocardiogram (ECG) beat classification [22], [23] and sleep stage classification using smaller epochs from a long Electroencephalogram (EEG) recordings [24], [25], with relatively less number of independent experiments or subjects, as in the present case as well.

In our previous work [17], analyses with different segment size of the plant signals (256, 512, 1024 samples) were carried out to tune the parameters of the drift removal high-pass filter. It is understandable that too small a segment size may result in the nonstationary signal behaving as a stationary signal [26], however most of the features of long range correlation like Hurst exponent and DFA will not be very meaningful in such a scenario. Also, within a very short time window, there may not be significant change in the statistical characteristics of the plant signal, as the inherent biological oscillations in plants are slow in nature. Therefore we have chosen the window heuristically as 1024 samples that allows taking sufficient samples for both short and long range feature extraction from these segmented time series.

The data segmentation is a valid approach with a stationary and ergodic assumption of the signal in both the pre and post-stimulus parts. A stationary ergodic process implies that the random process will not exhibit any significant change in its statistical properties with time and the statistical measures such as various moments (e.g. mean, variance etc.) of the stochastic process will be similar. As in the present scenario, where the plant electrical responses are typically non-stationary biological signals, we took two different approaches of pre-processing in order to transform them from non-stationary to stationary signals/features. This transformation assumes that the effect of the stimulus is embedded in the signals in a similar way between different segments. This also implies that the stimulus has modified one or more statistical features to a significant extent between the pre and post-stimulus parts whereas such variation between segments coming from the same pre or post-stimulus parts are not that significant. This criteria is enforced by the optimum high-pass IIR filter as designed in [17] since it brings the energy distribution of the background signals to a common reference point (centroid of the wavelet packet energies across different nodes).

The two pre-processing steps for enforcing this stationarity assumption is given below:

- When using raw plant signals for feature extraction followed by classification – the features were derived from small epochs of fixed sample size (1024 samples), from both pre and post stimulus parts of the signals. The mean values of each feature from the pre-stimuli parts, were then calculated and subtracted from the corresponding features of the post-stimulus parts. This ensured that the resultant pre-stimulus mean subtracted features of the post-stimulus parts reflect only the change in the statistical features due the introduction of the stimulus. Hence any bias due to the difference in the pre-stimulus signal's starting amplitude or slow time varying drift can be minimised. This approach has also been adopted in [2] for a similar classification problem.

- When using filtered plant signals, an optimum high-pass filter was applied on the raw signal to remove the low frequency trends and the remaining stochastic parts were used to divide the long signals in multiple epochs of fixed sample size (1024 samples). Features were calculated from each of these epochs from the post-stimulus parts of the signals since the high pass filter has already been tuned using the pre-stimulus information as reported in [17].

In both of these two methods, each segmented epochs were assumed to be independent and identically distributed (IID) belonging to the same class as of the original duration of the plant electrical signals. This segmentation may break the long range correlations in the data but resampling of long time series and using each segment as a different data point is a well-established method in other bio-electrical signal processing application as well e.g. ECG beat classification [22], [23] and dividing EEG signals into several epochs for sleep stage classification [24], [25], amongst many other similar time series analysis problems as discussed before. Resampling and windowing are a routine part of biomedical data (ECG/EEG) analysis like heart-beat segmentation and event related potential (ERP) epoch extraction from EEG and finally using them in classification. The field of plant electrophysiology based stimulus classification is new yet growing field and as such there is no benchmark exists for routine data analysis. We here propose data windowing and statistical feature extraction, as some standard pre-processing techniques to detect the stimulus to the plant.

## IV. Feature Extraction and Significance

In total, 15 features were extracted from each block containing 1024 samples. Out of these 15 features, 11 features were previously reported in [2] which includes basic measures of descriptive statistics like *mean* ($\mu$), *variance* ($\sigma^2$), *skewness* ($\gamma$), *kurtosis* ($\beta$), *Interquartile range* (*IQR*). It also includes features capturing nonlinear and non-stationary behaviour [27], e.g. *Hjorth mobility*, *Hjorth complexity*, *detrended fluctuation analysis* (*DFA*), *Hurst exponent*, *wavelet packet entropy* and also frequency domain features like the *average spectral power* etc. Apart from these 11, four additional features have been considered in the present paper. Two out of these four are higher standardized moments capturing the non-Gaussian nature of the signal given in (1).

$$S_{N=5,6} = \left( \left( x - \mu \right) / \sigma \right)^{N} \tag{1}$$

which is described as *hyperskewness* when $N = 5$ and *hyperflatness* when in $N = 6$ in (1) respectively [28]. The third additional feature is the *Fano factor* or the index of dispersion described as $F = \left( \sigma^2 / \mu \right)$ which also characterizes the shape of the underlying probability density function for the data generation process [29]. The fourth feature considered is the correlation dimension (*D*) which takes a fractional value for chaotic trajectories while taking integer values for regular objects like line or surface [30]. For large number of samples in a time series data, the correlation dimension (*D*) converges to the Hausdorff or fractal dimension ($D_f$). From a measured signal using the Takens' time delay embedding, the correlation sum is first computed and then inspected for self-similarity in a log-log plot and the slope of this graph gives the correlation dimension (*D*) [30].

These 15 features, extracted from blocks of 1024 samples, were normalized to lie within a minimum {0} and maximum {1} value by using (2).

$$\tilde{x} = \left( x - x_{\min} \right) / \left( x_{\max} - x_{\min} \right) \tag{2}$$

where, $x$ is the feature value, $x_{\min}$ and $x_{\max}$ are the minimum and maximum values of the feature vector respectively.

The same 15 features were also extracted from raw signals (background information subtracted) and normalized, so that we could compare the results obtained using both filtered and raw signals. The univariate histogram plots of the features, extracted from the filtered and raw data, are shown in Figure 5 and Figure 6 respectively to explore the class separation and overlaps using individual features. From these figures, we note that in most cases, the individual feature does not allow a straightforward separation of the classes.

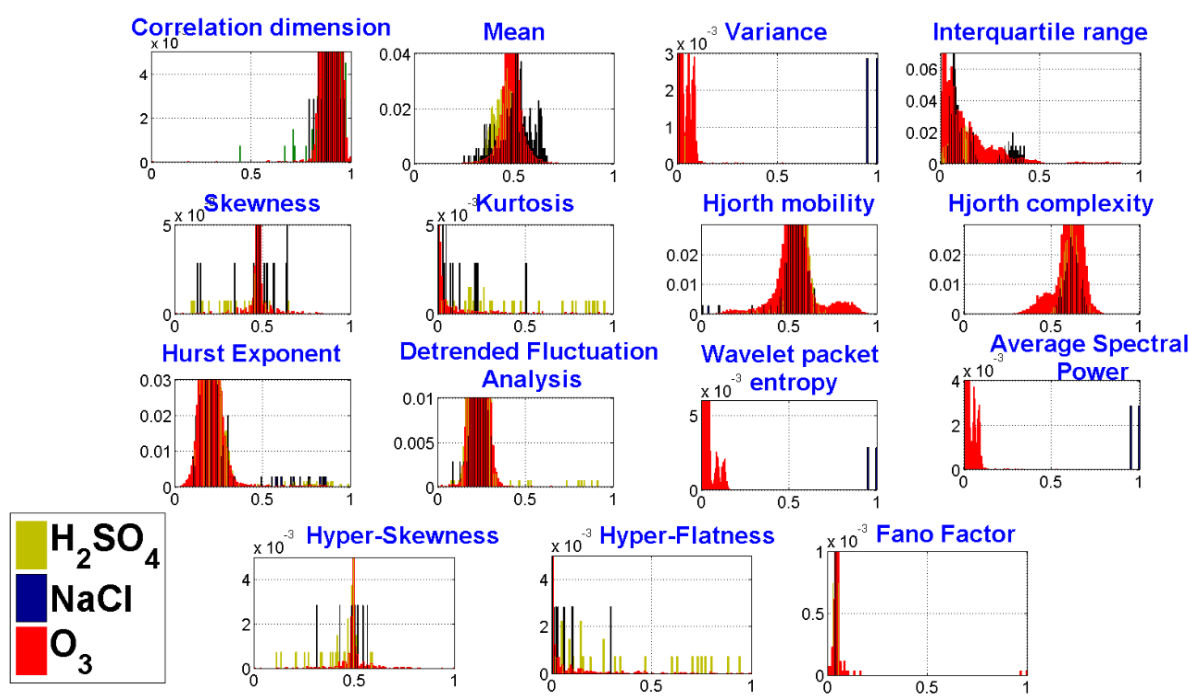

Figure 5: Histograms for 15 features (computed from filtered data), showing separation of different classes.

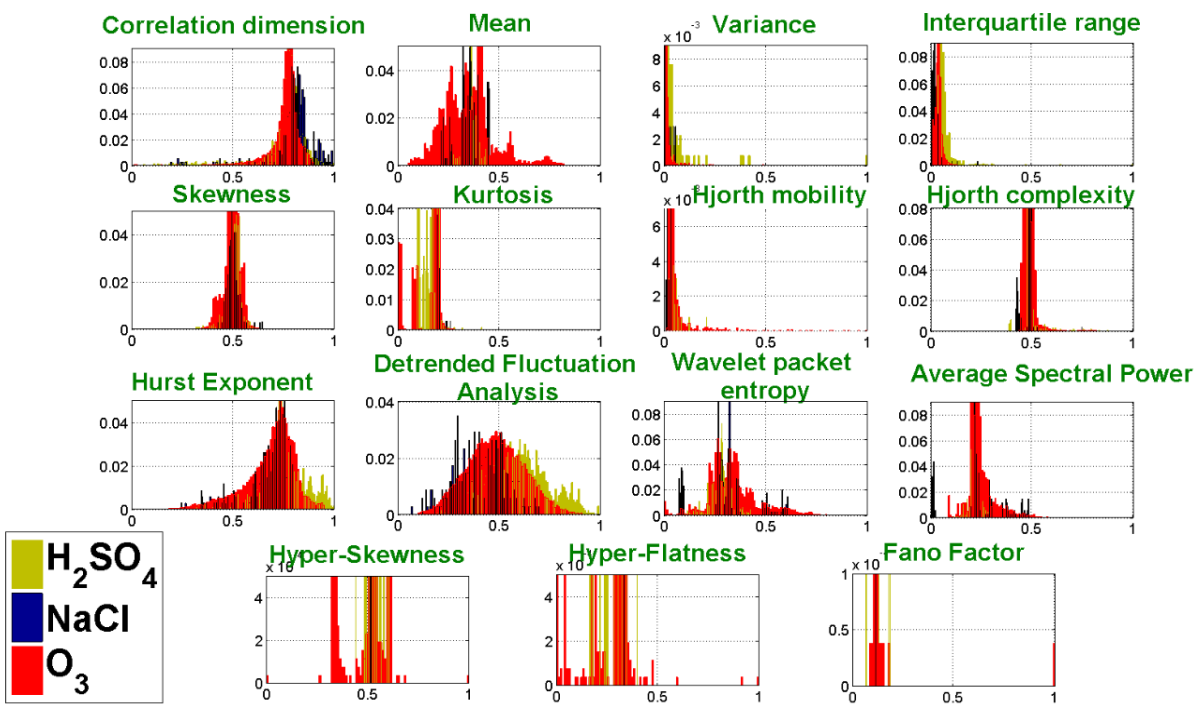

Figure 6: Histograms for 15 features (computed from raw data), showing separation of different classes.

It is worth mentioning that here the purpose of the study is not interpretation of these statistical features with changing biological behaviour of the plant under chemical stress. Rather the goal here is whether we can reliably detect the chemical stimuli by only observing these features. The classification results shown in the next sections indicate that the individual features alone may not be very useful to detect these changes in the underlying physiological characteristics of the plant under chemical stimulation and act as a clear biomarker as revealed by many overlapping univariate histograms in Figure 5 and Figure 6. But a particular combination of these features may indeed provide useful results to detect these external stimuli.

## V.    Classification Methodology

We have used four *discriminant analysis* classifier variants – LDA, QDA, diaglinear, diagquadratic and the *Mahalonobis distance* classification algorithms as also previously explored in [2], but in two different decision tree settings. The diaglinear and diagquadratic classifiers, which are also known as Naïve Bayes classifier, uses a simple linear and quadratic kernel along with only diagonal estimates of the covariance matrix (thereby neglecting any cross-terms or feature correlations). The Mahalonobis distance classifier modifies the distance measure instead of the standard Euclidean version [15], [16]. We have restricted the study using only these five classifier variants due to their low computational complexity on large datasets, compared to nonlinear kernel based classifiers like the support vector machines. This will assist in our final goal of implementing the best classification algorithm on an electronic embedded system within a resource constrained environment that can be deployed in real field.

### A.   Architecture of the Decision Tree

We explored two different types of decision tree architectures – OVR and OVO for the present multi-class classification problem involving three chemical stimuli $O_3$, NaCl and $H_2SO_4$. These are schematically shown in Figure 7 and Figure 8 respectively. In both the configurations, we explored the best feature-classifier combination which would produce the best accuracy in terms of both cross-validation and independent testing. Thus we have reduced a multiclass classification problem to one or more binary classification problems [31], [32] exploring two distinct decision tree methods, as described in the following sections using OVR [33] and OVO [34] schemes.

### B.   One *vs.* Rest (OVR) Scheme

In the OVR scheme [35], [36], the three classes (stimuli) were classified in a two-node set up. In the first node, the *best separable* class (out of the three) was found out (depicted as Class A in Figure 7) along with the best features-classifier combination. In the second node, the remaining two classes were classified. In our case, after the retrospective study, it was possible to associate the three chemical stimuli with the classes A, B and C, as shown in Figure 7.

In order to find out the best separable class out of the three classes, three binary classification settings were setup for the first node. In each binary setting, one of the stimuli was considered as *one* class and the other two were clubbed as the *rest*. For each of these settings, we could rank features using Fisher Discriminant Ratio (FDR) [15], [16] given by (3).

$$FDR = \left(\overline{\mu_1} - \overline{\mu_2}\right)^2 \Big/ \left(\overline{\sigma_1}^2 + \overline{\sigma_2}^2\right) \qquad (3)$$

where $\overline{\mu_1}$ and $\overline{\sigma_1}$ is mean and standard deviation of features of *one* class, $\overline{\mu_2}$ and $\overline{\sigma_2}$ are mean and standard deviation of features of the *rest*. This mean and standard deviation in (3) correspond to each feature for each class and therefore should not be confused with the signal means (raw or filtered), described before as potential features. A systematic feature selection method, employing Sequential Forward Search (SFS) algorithm [15], [16] on the FDR based ranked features, for each of the three binary classification settings (i.e. *Class A vs. rest*, *Class B vs. rest* and *Class C vs. rest*) was carried out using the five classifier variants. Thus the best feature-classifier combination was found out for the best separable class. The best feature-classifier combination for the remaining two classes were also found out using a similar approach (i.e. ranking the features using FDR, then employing SFS along with five different classifiers).

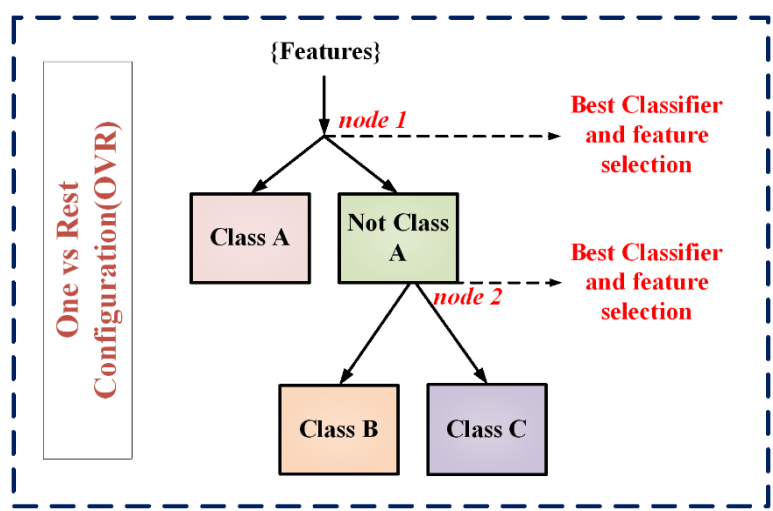

Figure 7: Decision tree incorporating OVR configuration for multi-class classification.

## C. One *vs.* One (OVO) Scheme

In the case of OVO configuration, three binary classification settings were simultaneously carried out (as shown in Figure 8). If two classifiers affirm the presence of a particular class, then only that particular class was predicted [37]. In the case of contradictory decisions by two classifiers, such an assignment was discarded. For each binary classification settings, the best feature-classifier combination was found out. The feature ranking for each of the three settings were again recalculated using FDR, as explained earlier for different pairs of classes. SFS was employed again on these FDR based ranked features for all the five classifier variants to find out the best feature-classifier combination for each of the three binary classifiers shown in Figure 8.

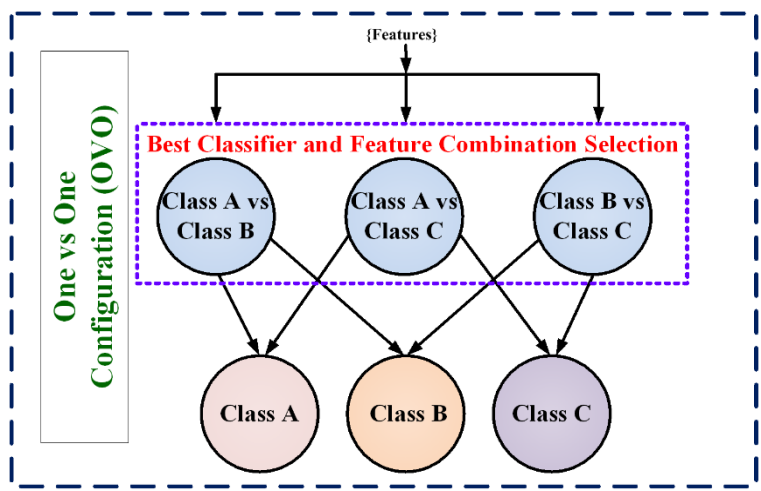

Figure 8: Decision tree incorporating OVO configuration for multi-class classification.

### D. **Retrospective Study (Using LOOCV)**

The results for identifying each class in a binary classification are obtained in terms of sensitivity and specificity given by the confusion matrix in Table 3 [15], [16].

Table 3: Confusion matrix

| Predicted class | Actual class | |
|---|---|---|
| | **Positive (P)** | **Negative (N)** |
| **Positive (p)** | True Positive (TP) | False Positive (FP) |
| **Negative (n)** | False Negative (FN) | True Negative (TN) |
| **Measures** | *Sensitivity*=TP/(TP+FN) | *Specificity*=TN/(TN+FP) |

For cases, where $P \approx N$, accuracy of the classification is obtained using the traditional notion of accuracy in (4).

$$Accuracy = \left( TP + TN \right) \big/ \left( P + N \right) \tag{4}$$

But in the case of unbalanced data [38] in the two classes (i.e. $P \gg N$ or vice versa), the *balanced accuracy* [39]–[42] is used to determine the accuracy of the classification as given by (5). The derivation of (5) is given in [38] for a single run of the classifier and for a fixed threshold in the classifier's class assignment.

$$Accuracy(balanced) = \left( sensitivity + specificity \right) \big/ 2 \tag{5}$$

Since we have an unbalanced dataset due to different duration of exposure of the plants to different stimuli, the *balanced accuracy* has been used here which henceforth will be referred as *accuracy*, in the remainder of this paper. Identifying the difficult class within an unbalanced data-set has been the focus of several recent works like [43], as also attempted here.

We thus explored the best combination of features and classifiers to get the optimum classification results within the decision tree framework in the following way:

- The features for every binary classification setting (OVR and OVO) were ranked using FDR as given by (3). Depending on the OVR/OVO scheme, the feature ranking may get changed even though the feature descriptions are constant.

- Using the ranked features in each binary classification setting, a SFS algorithm was employed with all the five classifiers to find the best result.

The ranked features for each binary classification settings are provided in the supplementary material, for brevity.

### E. **Visualization of Class Separability on the LDA Basis**

The LDA basis (*Fisher-faces*) [15], [16] vectors are found by solving the generalized eigenvalue problem as given by (6).

$$S_w^{-1} S_b W = \lambda W \qquad (6)$$

The terms $S_b$ and $S_W$ are called between-class and within-class scatter matrix respectively for the original *n*-dimensional feature vector extracted from the plant signals.

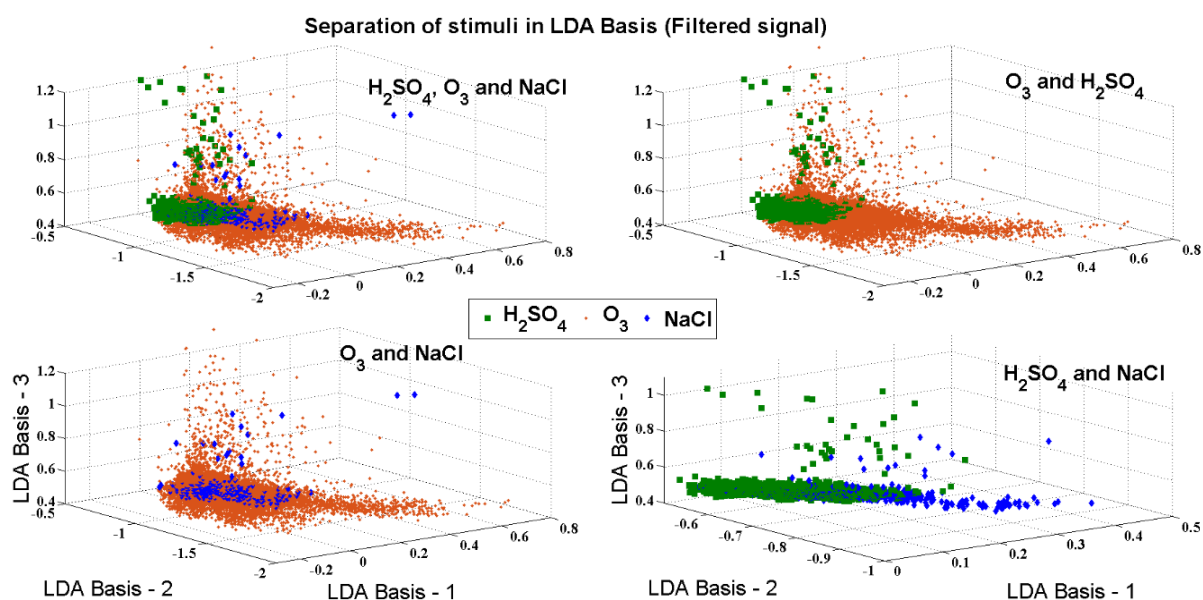

Figure 9: LDA basis showing separation of three stimuli using filtered signal.

The eigenvectors corresponding to the largest eigenvalue gives the optimal projection where the variance between the features within a class is maximized. The three dimensions of the LDA basis are obtained from eigenvectors corresponding to the three largest eigenvalues. The separability of the three stimuli on 3-dimensional LDA basis using filtered and raw signals are shown in Figure 9 and Figure 10 respectively. The data shown in Figure 9 and Figure 10 are those which are used for training (~73% of total data) the classifiers during LOOCV retrospective study.

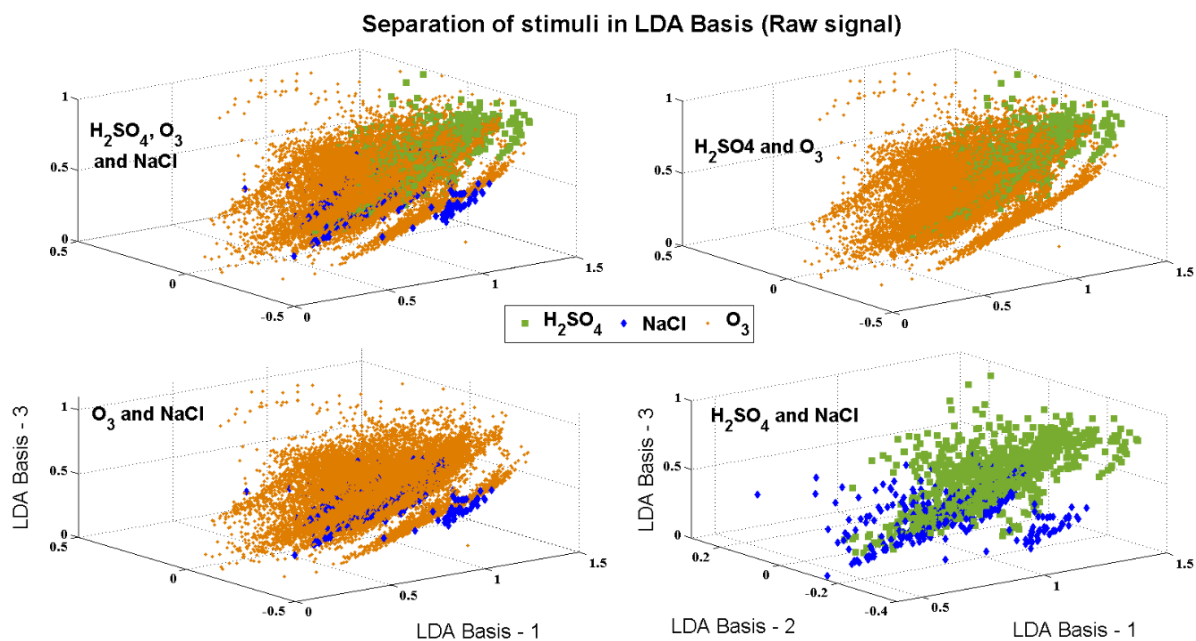

Figure 10: LDA basis showing separation of three stimuli using raw signal.

# VI. Results and Discussions

## A. Retrospective Study Using Filtered Plant Signals

We here report the results obtained using the filtered plant signals. The classification accuracy obtained using the SFS algorithm on the ranked features and using all the five variants of classifiers are shown in Figure 11 (OVR) and Figure 12 (OVO), from which the classifier and feature combination giving the best accuracy could be easily identified (as the maxima of the curves) for each binary classification scenario. These results are obtained by carrying out *LOOCV* on the ~73% training dataset for the retrospective study as explained earlier. In Table 4, along with the best results in terms of percentage of accuracy, the top features and classifiers have also been reported. As can be seen from Table 4, the results are obtained for different binary classification scenario. For every such scenario, the features have been ranked using FDR given in (3), and hence not all the cases will have the same features at the same ranking.

From Table 4 we can observe that 100% classification is achieved between NaCl and $H_2SO_4$ using top 6 features and QDA/Mahalanobis classifier. This result is quite significant since both the stimuli are administered through the soil (and hence the uptake is through the roots). A similar result is obtained for NaCl and $O_3$ although using top 15 features. Here $O_3$ is administered through a spray all over the plant, so the absorption is mainly through the leaves. The requirement of 15 dimensional feature space to distinguish between the two stimuli probably means the effect of how the stimuli is applied to the plant is irrelevant, i.e. whether through the root or leaf. However, perhaps it also shows that the electrophysiological effect in the plant due to NaCl is different to a great extent than the other two stimuli as it produces a 100% classification result.

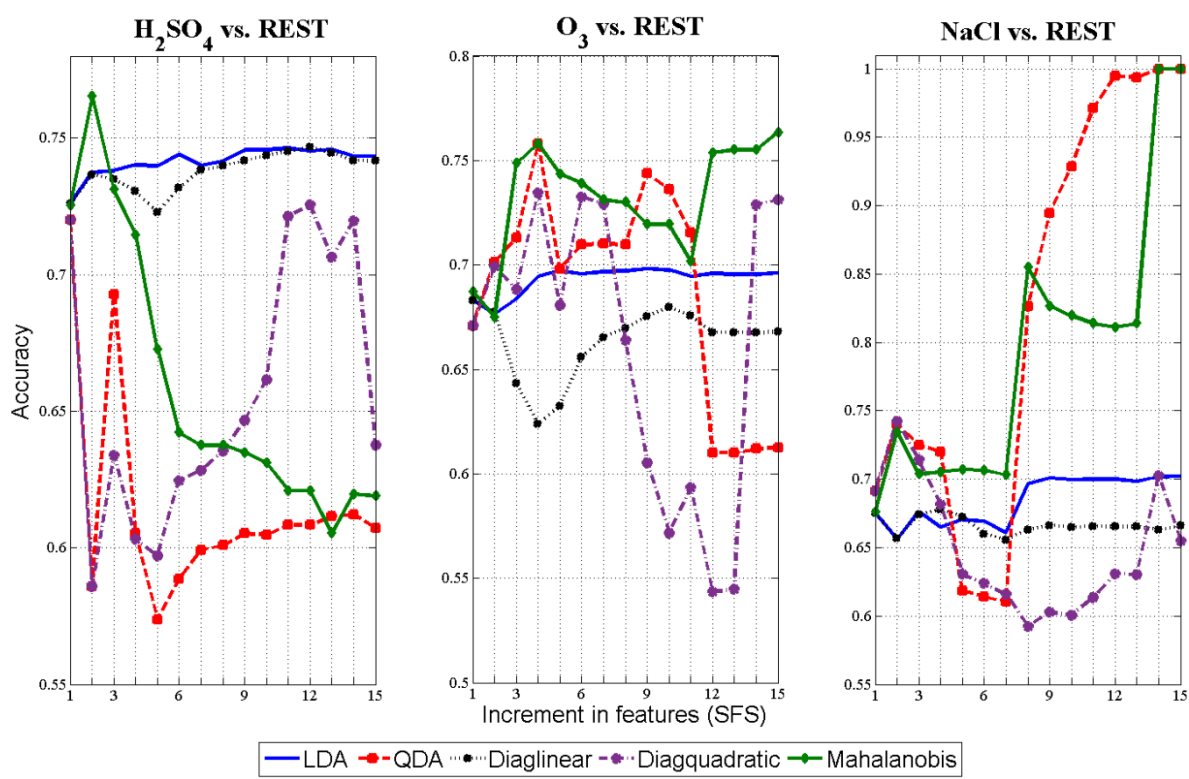

Figure 11: Accuracy *vs.* increment in features (SFS) for OVR setting (using filtered signal).

Table 4: Best classification accuracy for the filtered signal (features + classifier combinations)

| Scheme | OVO | | OVR |
|---|---|---|---|
| Stimuli | NaCl | $O_3$ | Rest |
| $H_2SO_4$ | 100% (top 6 features), *QDA / Mahalanobis* | 76.53% (top 2 features), *Mahalanobis* | 76.53% (top 2 features), *Mahalanobis* |
| NaCl | - | 100% (top 15 features), *QDA / Mahalanobis* | 100% (top 14 features), *QDA / Mahalanobis* |
| $O_3$ | * | - | 76.34% (top 15 features), *Mahalanobis* |

The best result between $O_3$ and $H_2SO_4$ is approximately 76% which is not as good as that obtained using NaCl stimulus. This observation possibly points to the fact that the electrophysiological effect due to $O_3$ is not as separable from the effect of $H_2SO_4$, especially using low dimensional discriminant analysis classifiers.

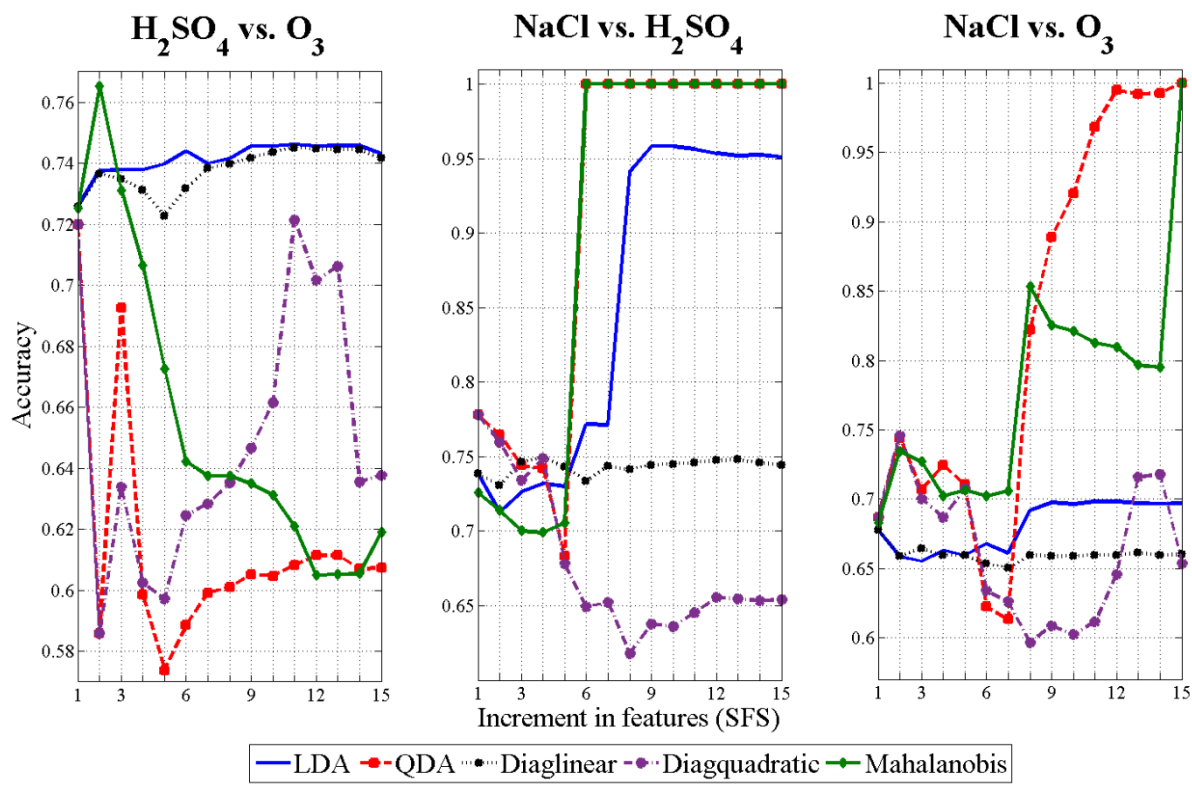

Figure 12: Accuracy *vs.* increment in features (SFS) for OVO setting (using filtered signal).

*B.* **Constructing the Decision Tree for Multiple Class Using Binary Classification Results of the Filtered Plant Signals**

The decision tree was constructed using two configurations – OVR and OVO, as explained in earlier sections. From Table 4, we can see that *NaCl vs. Rest* gives the best classification result of 100%, using top 14 features (as ranked using FDR) along with QDA/Mahalanobis distance classifier. So in the first node of Figure 7, we set class A = NaCl as the best separable stimuli from the rest. Thus the decision tree will first test if an incoming feature vector belongs to NaCl or not. If it is tested to be *not NaCl*, then the remaining binary classification to be evaluated in the second node is between *$O_3$ vs. $H_2SO_4$* (= class B and class C respectively in Figure 7). As seen from Table 4, the best result achieved for classification between $O_3$ and $H_2SO_4$ is 76.53% using top two features along with the Mahalanobis distance classifier.

Thus, following the retrospective study we found out that the decision tree in OVR configuration will require top 14 features (*NaCl vs. Rest*) along with QDA/Mahalanobis classifier in the first node to test an incoming feature vector to be classified as NaCl. If it is found to be not from NaCl, then in the second node, the feature vector is tested for $O_3$ or $H_2SO_4$ using top two features and Mahalanobis classifier.

In the OVO configuration, three binary classifier settings were used. As can be seen from Table 4, *NaCl vs. $O_3$* and *NaCl vs. $H_2SO_4$* achieve 100% accuracy using top six and top 15 features respectively, along with either QDA or Mahalanobis distance classifier. The *$O_3$ vs. $H_2SO_4$* achieves the best result of 76.53% using top two features along with Mahalanobis distance classifier.

Therefore after the retrospective study, we see that the decision tree in OVO configuration will test an incoming feature vector for three binary classification settings simultaneously. The classifier setting of *NaCl vs. $O_3$* will require top six features along with QDA/Mahalanobis classifier. The next classifier setting of *NaCl vs. $H_2SO_4$* will require top 15 features along with QDA/Mahalanobis classifier. The last classifier setting of *$O_3$ vs. $H_2SO_4$* will require top two features along with Mahalanobis classifier.

*C.* **Prospective Study Using Features from the Filtered Signal**

One experimental dataset from each stimuli were set aside for making up the independent test dataset for the prospective study. The results, which were obtained using the retrospective study with LOOCV, were then used to design the decision tree (in OVR/OVO configuration). These two decision trees were then used to test the streams of features extracted from the signals constituting the independent test dataset. These features, as mentioned in section III, were extracted from a block of 1024 samples from the post-stimulus section of the signals belonging to the test dataset.

For each of the two decision tree configurations – OVR and OVO, a decision criteria based on the value of the classifier function interpreted which stimuli the features point towards. The answers were then measured against the known class labels of the test dataset, thereby determining the accuracy of the configuration. During the prospective study, if the decision tree could not give a definite class for the input feature vector (computed from the independent test data), then the resulting class was termed as *Unknown*. The results for each of the three stimuli were collected according to (7).

$$Accuracy_{prospective} = \left( c_a + c_b + c_c \right) / \left( n_a + n_b + n_c \right) \tag{7}$$

Here $\{c_a, c_b, c_c\}$ denotes the correct number of blocks belonging to each of the three classes (stimuli) detected by the decision tree. The total number of blocks belonging to the three stimuli are denoted by $\{n_a, n_b, n_c\}$ and can be referred from Table 2. From the results of retrospective study in Table 4, we found out how many ranked features were required for the best results for classification of each stimuli within the training dataset (with LOOCV). Here it needs to be noted that the feature and classifier settings achieved through retrospective study, did not yield similar good results during prospective study i.e. in OVR configuration, using top 14 features to detect NaCl in the first node and top 2 features to detect $O_3$ / $H_2SO_4$ in the second node along with the Mahalanobis distance classifier (in both nodes) managed to produce an accuracy of only 33%. It failed to detect any blocks from $H_2SO_4$ and only one block from $O_3$. However changing the classifier to LDA in the first node (keeping Mahalanobis in the second node) improved the accuracy to around 74% through increased detection of $H_2SO_4$ (134 out of 148 blocks) and $O_3$ (5862 out of 9692 blocks). However from Figure 11, we clearly see that LDA in the first node produces an accuracy of less than 70% for *NaCl vs. Rest*. A similar poor result was obtained during prospective study when using OVO configuration. Thus we decided to check the effect of increasing features (using SFS algorithm) in each node simultaneously to see which features and classifier combination produces good results for both retrospective and prospective studies.

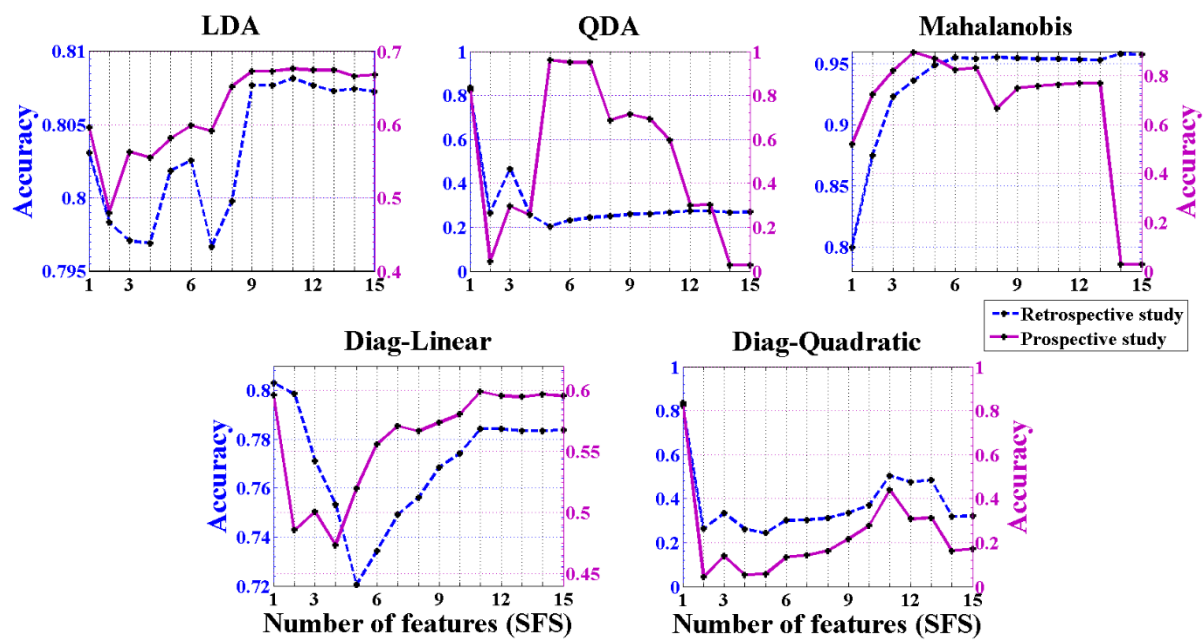

Figure 13: Retrospective *vs.* prospective study results for OVR configuration, using five different classifiers and SFS.

To achieve a good classification accuracy for both training and testing datasets, we compared the results of correctly detected blocks from each stimulus for both retrospective and prospective study by incrementing the ranked features using SFS. In other words, for each of the two classifiers (at two nodes) in OVR, we increased the features simultaneously. This is best understood by referring back to Figure 7, where we now know that at node 1, the binary classification to be tested is *NaCl vs. rest* and at node 2 it is $H_2SO_4$ *vs.* $O_3$ (as established through retrospective study). So we started with best feature in both the nodes and also noted the number of correctly classified blocks from NaCl, $H_2SO_4$ and $O_3$. Then we incremented the number of features by one and noted the results and it continued till all the 15 features were employed in SFS. This process was repeated for all the five variants of the discriminant analysis classifiers. Thus we now have classification results involving independent test data, for incrementing features and five different classifiers.

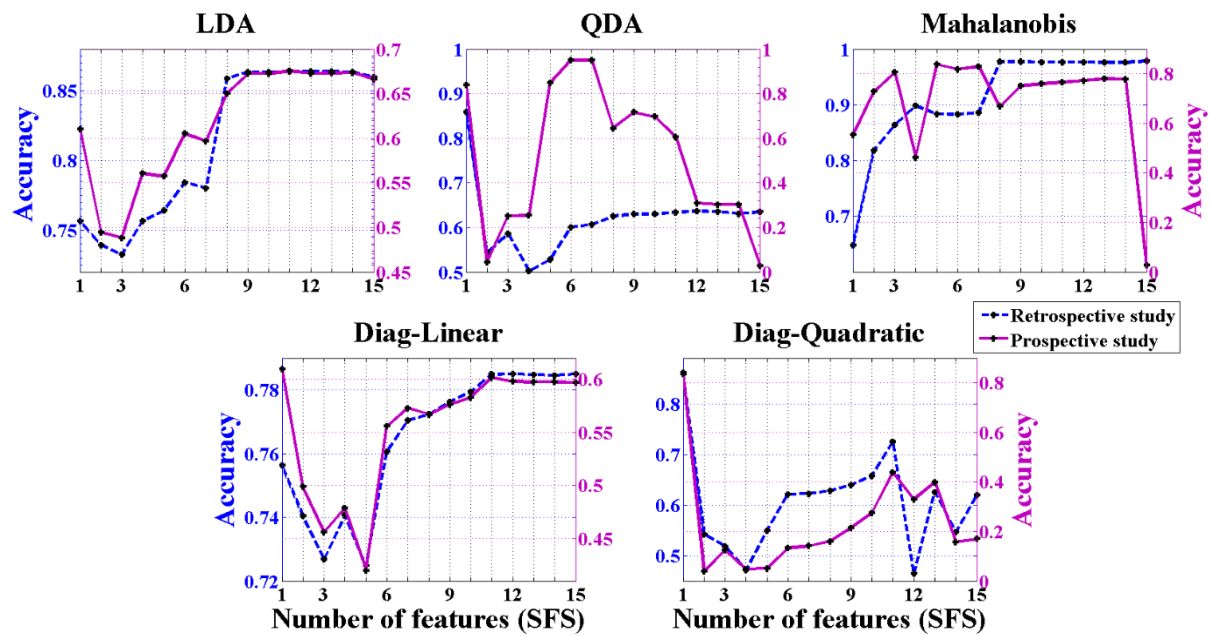

Figure 14: Retrospective *vs.* prospective study results for OVO configuration, using five different classifiers and SFS.

Similarly for OVO, we started with top one feature for all three classifiers (for three different binary classification) and noted the correctly classified blocks from each of the three stimuli. Then we incremented the feature by one and noted the results and it continued till all the features were employed in the SFS. Again, this process was repeated for all the five classifiers. We also had results from the retrospective study where the correctly classified block from each stimulus was noted for every increment in feature in the SFS for each of the five discriminant classifiers.

From the results obtained for both retrospective and prospective study in OVR configuration as shown in Figure 13, we could clearly observe that using top three features through top seven features and Mahalanobis classifier produced the best classification results of greater than 82% in the prospective study. Similarly, using top three and above features along with Mahalanobis classifier produces a classification result of greater than 92% in the retrospective study.

Among the two naïve Bayes classifiers, diagquadratic classifier performed well using top one feature providing accuracy greater than 82% in both retrospective and prospective study whereas diaglinear classifier provided a mediocre performance of around 59% during the prospective study. Among other classifiers, QDA provided an accuracy of greater than 82% in retrospective and 83% in prospective study by using top one feature, whereas LDA classifier produced a classification accuracy of greater than 80% and 59% in retrospective and prospective study respectively.

Next, we looked at the results from OVO configuration as shown in Figure 14. The best results of classification accuracy were greater than 88% in retrospective and greater than 83% in prospective study and was obtained using top five features and Mahalanobis classifier. The LDA produced the best classification accuracy of ~86% during retrospective study and ~67% during prospective study using top eleven features. The QDA on the other hand produced a classification accuracy of ~85% and ~83% during retrospective and prospective studies respectively, using the first ranked feature. Among the naïve Bayes classifiers, diagquadratic classifier again performed well using top one feature providing accuracy of around 85% during retrospective and ~83% during prospective studies whereas diaglinear classifier provided a mediocre performance of around 60% during prospective study and somewhat acceptable performance of ~75% during retrospective study.

As a summary, we can conclude that using top four features computed from the stochastic part of plant electrical response, along with Mahalanobis distance classifier, provides a best multiclass classification accuracy of ~93% in retrospective and ~89% in prospective study in an OVR setting. The detailed list of features has been provided in the supplementary material for convenience. In the next subsection, we compared these results with those obtained using features from the raw signals (with background information subtracted) as also explored in [2]. Here we used an extended feature set and more exhaustive experimental data for calculating both independent and cross-validation accuracy.

*D.* **Retrospective Study Using the Raw Plant Signals**
We adopted the same methodology for classification using features from raw signals as we did for the filtered plant signal. The classification accuracy obtained using the SFS algorithm on ranked features and using all the five variants of discriminant analysis classifiers are shown in Figure 15 (OVR) and in Figure 16 (OVO). From these, we identified the classifier and feature combination giving the best accuracy for each binary classification scenario. These results were obtained by carrying out *LOOCV* on the ~73% training dataset for the retrospective study as explained earlier. The feature ranks for each binary classification scenario is given in the supplementary material.

The best results obtained in retrospective study (~73% dataset, LOOCV) in OVR/OVO setting is given in Table 5 from which we found out that Class {A = NaCl, B = $H_2SO_4$, C = $O_3$} respectively in the OVR setting. Comparing these results with those obtained with the filtered signals, we observe that any binary classification results involving NaCl performs much better than other two chemical stimuli. As can be compared from Table 4 and Table 5, 100% accuracy is achieved for any binary classification involving NaCl when using filtered signals compared to ~90% achieved using raw signals. Classification result for $O_3$ and $H_2SO_4$ using raw signal during retrospective study is around ~78% which is marginally higher than that obtained using the filtered signal.

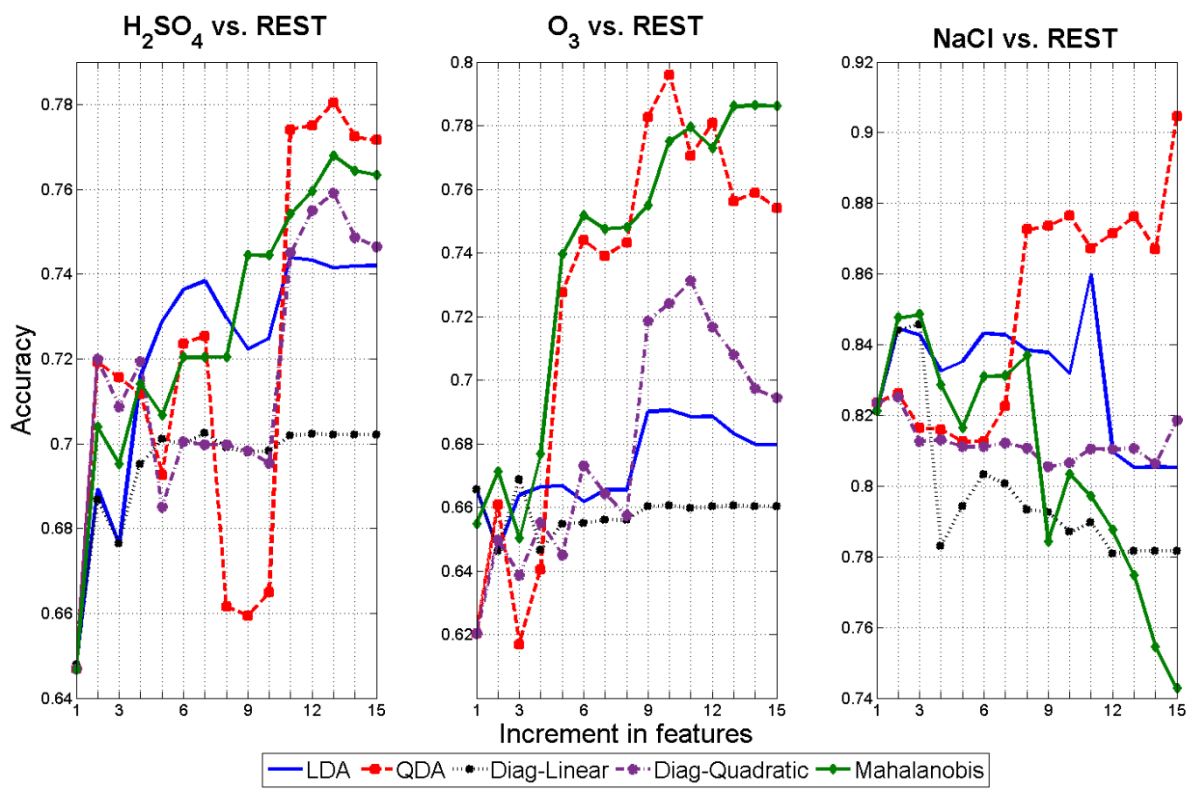

Figure 15: Accuracy *vs.* increment in features (SFS) for OVR setting (using raw signal).

Table 5: Best classification accuracy for the raw signal (features + classifier combinations).

| Scheme | OVO | | OVR |
|---|---|---|---|
| Stimuli | NaCl | $O_3$ | Rest |
| $H_2SO_4$ | 89.43% (top 14 features), *LDA* | 78.11% (top 13 features), *QDA* | 78.04% (top 13 features), *QDA* |
| NaCl | - | 90.85% (top 15 features), *LDA* | 90.47% (top 15 features), *QDA* |
| Ozone | * | - | 79.58% (top 10 features), *QDA* |

Also in the case of filtered signals, a higher dimensional feature space was required for achieving the best classification accuracy for *NaCl vs. Ozone* (15 features), *NaCl vs. rest* (14 features) and *Ozone vs. rest* (15 features). Rest of the binary combinations required comparatively lower dimensional feature space. When this was compared to the classification results using raw signals, we saw that for discriminating *Ozone vs. rest* we required top 10 features for achieving the best classification results. All other binary classification combinations required 13 or more features. Therefore the separability of any two stimuli in the feature space is comparatively better for filtered signal than the raw signals.

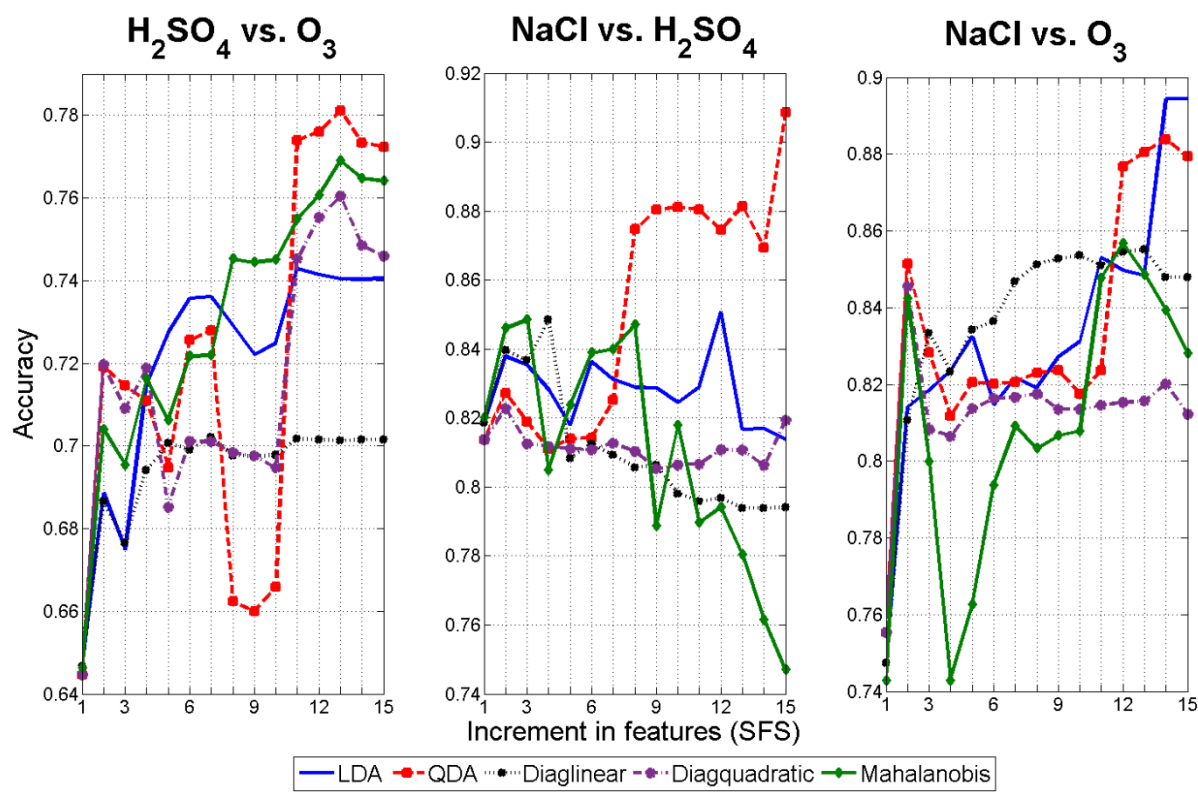

Figure 16: Accuracy *vs.* increment in features (SFS) for OVO setting (using raw signal).

*E.* **Constructing the Decision Tree for Multiple Classes Using Binary Classification Results from the Raw Plant Signals**

Thus, following the retrospective study we found out that the decision tree in OVR configuration using raw signals will require top 15 features (*NaCl vs. Rest*) along with QDA classifier in the first node to test an incoming feature vector for belonging to NaCl class. If it is found to be not from NaCl, then in the second node, the feature vector is tested for belonging to either $O_3$ or $H_2SO_4$ using top 13 features and QDA classifier.

In OVO configuration, three binary classifier settings have been used as explained before for the filtered signals. As can be seen from Table 5, *NaCl vs. $O_3$* and *NaCl vs. $H_2SO_4$* achieve ~90% and ~89% accuracy using top 15 and top 14 features respectively, along with the LDA classifier. The *$O_3$ vs. $H_2SO_4$* achieves the best result of 78.11% using top 13 features along with QDA classifier.

Therefore after the retrospective study, we noted that the decision tree in OVO configuration using raw plant electrical signals will test an incoming feature vector for three binary classification settings simultaneously. The classifier setting of *NaCl vs. $O_3$* will require top 15 features along with LDA classifier. The next classifier setting of *NaCl vs. $H_2SO_4$* will require top 14 features along with LDA classifier. The last classifier setting of $O_3$ vs. $H_2SO_4$ will require top 13 features along with QDA classifier.

### F. Prospective Study Using Features from Raw Data

However, when the configurations obtained from retrospective study were used during prospective study, classification accuracy was as low as 30% which was similar to as found when using filtered signals. The results of prospective and retrospective study with respect to increment in ranked features (using SFS) obtained through five different classifiers are shown in Figure 17.

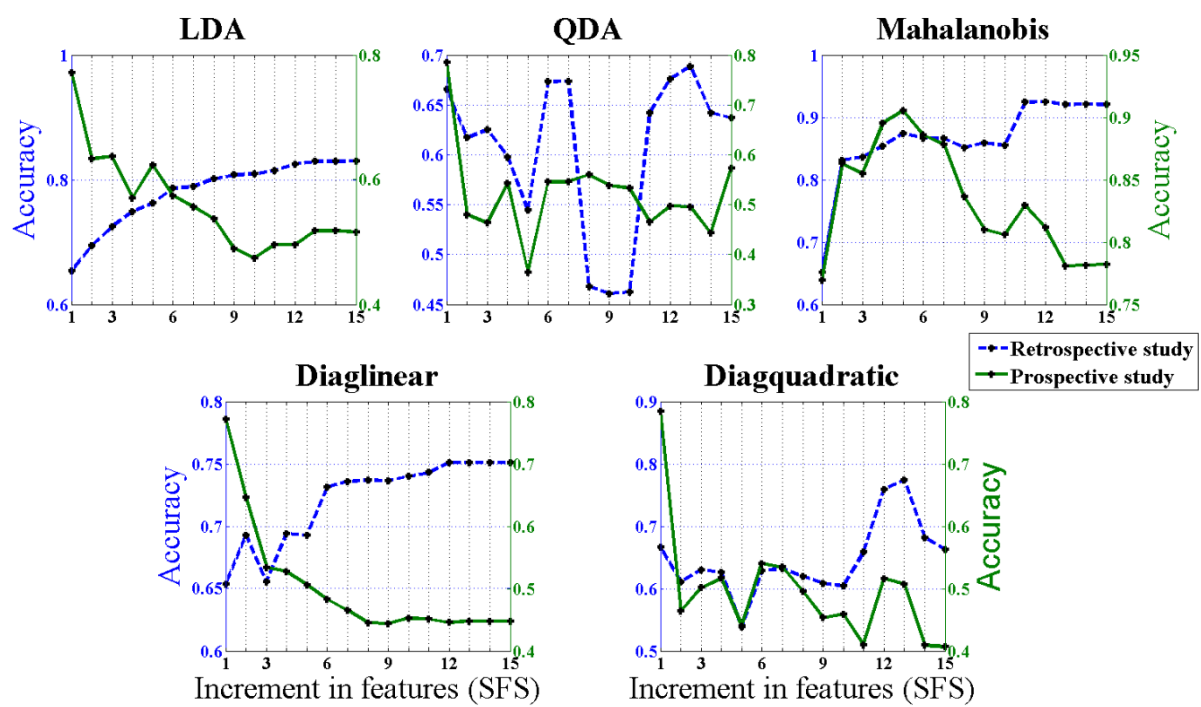

Figure 17: Retrospective *vs.* prospective study results for OVR configuration, using five different classifiers and SFS (features computed from raw signals).

Here again it needs to be noted that we checked the effect of incrementing ranked features (using SFS algorithm) in order to see which features and classifier combination produced good results for both retrospective and prospective studies rather than using those classifier-feature combinations which yielded the best results only during the retrospective study. The reason behind this approach is that using different partitions of the data may result in inconsistent results between the prospective and retrospective analysis, particularly using higher number of features and more complex classifiers. However a good practice should include keeping the number of features to be low and verifying the accuracy of the classifier on the held-out data based prospective study while using the best classifier-feature settings obtained from the retrospective study with LOOCV.

We observed that Mahalanobis classifier produced the best results for both retrospective (~87%) and prospective (~90%) study in an OVR setting. Although the prospective results were around 77% when using the naïve Bayes (diaglinear and diagquadratic) classifiers along with the top 1 feature, the retrospective results were limited to below ~70%. The LDA and QDA also perform mediocre as can be seen from Figure 17.

Looking at the OVO settings as shown in Figure 18, we conclude that using the top five features computed from raw plant electrical signals, Mahalanobis classifier produces the best results for both retrospective (~92%) and prospective (~90%) study in an OVO setting. LDA, QDA and the naïve Bayes (diaglinear and diagquadratic) classifiers produce above 75% results using top one feature for both retrospective and prospective study. Increasing the features produce good classification results for retrospective study but deteriorates the performance for the prospective study.

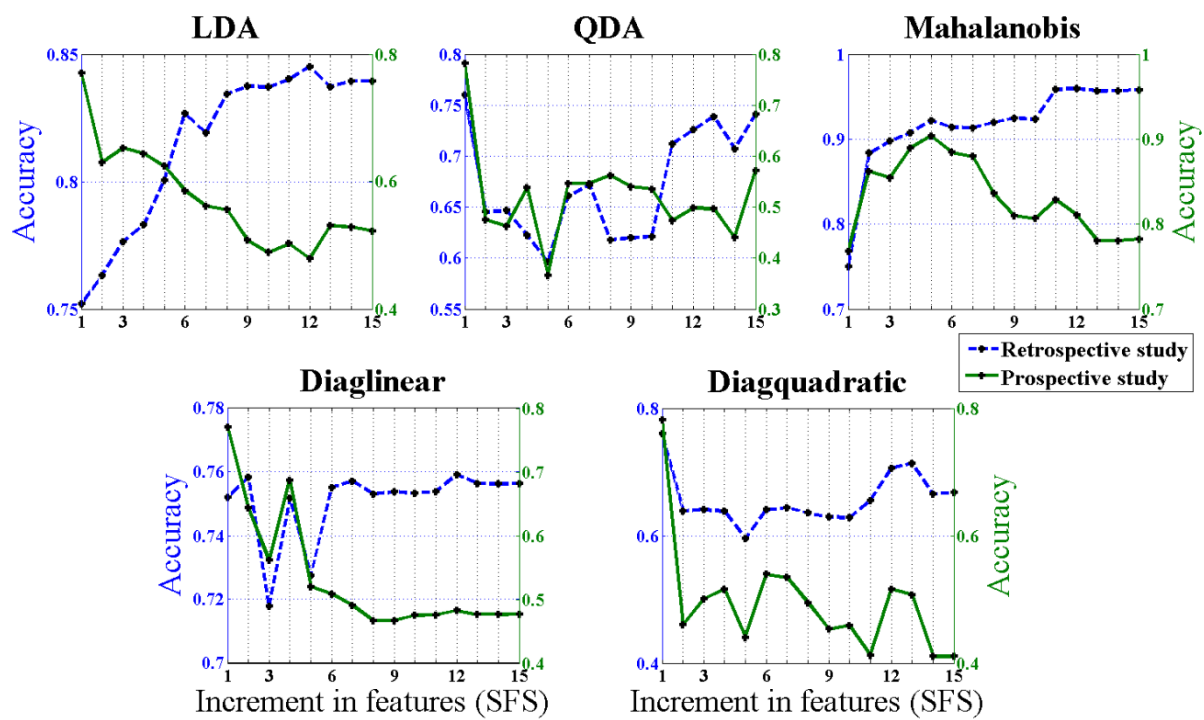

Figure 18: Retrospective *vs.* prospective study results (raw signal with background information subtracted) for OVO configuration, using five different classifiers and SFS.

From the above exploration the following recommendations could be made:

- The Mahalanobis distance classifier performs the best for both raw and filtered signals.

- However when compared between raw and filtered signals, the results are better for raw signals in OVO settings using the top five features.

- These top five features, as can be found in the supplementary material, are different for different binary combinations and include *IQR*, *hyper-flatness*, *kurtosis*, *variance*, *DFA*, *hyper-skewness*, *wavelet entropy* and *average spectral power*.

Therefore the recommended feature-classifier combinations and the decision tree architecture, presented in this paper could be considered as a thorough exploration to classify three externally applied chemical stimuli using plant electrical signals. The optimal architecture balances both the cross-validated training and independent held-out testing accuracies of the classifier tree. In addition, a reduced number of features have been recommended here even though it slightly sacrifices the classification accuracy in order to facilitate electronic implementation of the classification algorithm in a resource constrained environment in a future study [44].

Regarding the adopted data partitioning approach to test various classification methods on limited number of data-points, as such there is no consensus in the contemporary literature and many different approaches have been proposed. As a summary, Molinaro *et al.* [45] have provided a detailed comparison of different resampling methods for classification tasks including LOOCV, N-fold cross validation ($N$ = 2, 5, 10), Monte Carlo cross validation, 0.632 bootstrap with and without replacement, split or hold out (1/2, 1/3 partition) methods. In this study, we first split the data in training with LOOCV (73%) and independent hold out testing (27%) which is in agreement with the recommendations of [45] and also enjoys the benefit of an acceptable bias-variance trade-off, although our method increases the computational burden due to multiple LOOCV runs in each node of the decision tree. A less computationally expensive option at the cost of less accurate solution can be using combination of $N$-fold cross validation and partitioning/split methods. Here we not only select the best models based on LOOCV alone but also use independent held out validation results which makes our findings even more substantial.

## VII.    Conclusion

The paper reports an exhaustive exploration of designing a decision tree classifier based on five different discriminant analysis classifier and 15 statistical features extracted from plant electrical signals. We employ two multiclass classification strategies OVO and OVR along with retrospective and prospective testing of the classifier to establish its generalization capability. It is found that amongst the three chemical stimuli - NaCl, in general is best separable compared to $O_3$ and $H_2SO_4$. Future scope of work could be implementing the decision tree algorithm along with the statistical features in an electronic hardware platform to develop a plant based novel biosensor that the EU FP7 project PLEASED aims to develop.

We also acknowledge that in this paper, we did not explicitly consider the underlying mechanism of plant physiology for understanding how different signals are generated by plants as an effect of applying different chemical stimuli. In other words, we here aim to develop a mapping in which we simply observe the electrical signals generated by the plants when subjected to specific chemical stimuli under controlled laboratory conditions. Considering some elements of the underlying plant electrophysiological knowledge in future works may provide deeper insights to implement perhaps more effective classification strategies.

## ACKNOWLEDGMENTS

The work reported in this paper was supported by project PLants Employed As SEnsor Devices (PLEASED), EC grant agreement number 296582. The experimental data are available in the PLEASED website at http://pleased-fp7.eu/?page_id=253.